\documentclass[11pt,letterpaper]{article}

\usepackage[numbers]{natbib}

\pdfoutput=1

\usepackage{color}
\usepackage[table]{xcolor}
\usepackage{dcolumn}
\usepackage{caption}
\usepackage{graphicx}

\usepackage[T1]{fontenc}
\usepackage[utf8]{inputenc}
\usepackage{lmodern}
\usepackage[font=small]{caption}
\usepackage{varwidth}

\usepackage[activate={true,nocompatibility},final,tracking=true,kerning=true,spacing=true,factor=1100,stretch=10,shrink=10]{microtype}
\microtypecontext{spacing=nonfrench}
\usepackage{xspace}

\usepackage{rotating}
\usepackage{caption,subcaption}
\usepackage{float}
\usepackage{psfrag}
\usepackage{tabularx}
\usepackage[hyphens]{url}
\usepackage{wrapfig}
\usepackage{longtable}
\usepackage{verbatim}
\usepackage{array,booktabs,multicol,multirow}

\usepackage[english]{babel}
\usepackage{textcomp}
\usepackage{csquotes}

\usepackage{footnote}
\makesavenoteenv{tabular}
\makesavenoteenv{table}

\usepackage{placeins}

\usepackage{makecell}

\usepackage{enumitem}
\setlist{leftmargin=7mm}

\makeatletter
\newcommand\footnoteref[1]{\protected@xdef\@thefnmark{\ref{#1}}\@footnotemark}
\makeatother

\usepackage[bookmarks, bookmarksopen, bookmarksnumbered, colorlinks,linkcolor=blue,urlcolor=blue,citecolor=blue]{hyperref}
\usepackage[all]{hypcap}
\definecolor{orange}{rgb}{1.0,0.3,0.0}
\definecolor{violet}{rgb}{0.75,0,1}
\definecolor{darkgreen}{rgb}{0,0.6,0}
\definecolor{cyan}{rgb}{0.2,0.7,0.7}
\definecolor{blueish}{rgb}{0.2,0.2,0.8}
\definecolor{darkblue}{rgb}{0.1,0.1,0.9}
\definecolor{lightgray}{gray}{0.9}

\newcommand*\rot{\rotatebox{90}}

\sloppypar

\begin{document}

\title{The State of Sustainable Research Software:
Results from the Workshop on Sustainable Software for Science: Practice
and Experiences (WSSSPE5.1)}
\author{Daniel S.~Katz\thanks{National Center for Supercomputing Applications (NCSA) \& Department of Computer Science  \& Department of Electrical and Computer Engineering  \& School of Information Sciences (iSchool), University of Illinois, Urbana--Champaign, IL, USA; d.katz@ieee.org; ORCID: 0000-0001-5934-7525},
Stephan Druskat\thanks{Department of German Studies and Linguistics, Humboldt-Universität zu Berlin, Berlin, Germany; stephan.druskat@hu-berlin.de; ORCID: 0000-0003-4925-7248},
Robert Haines\thanks{Research IT, University of Manchester, Manchester, UK; robert.haines@manchester.ac.uk; ORCID: 0000-0002-9538-7919},\\
Caroline Jay\thanks{School of Computer Science, University of Manchester, Manchester, UK; caroline.jay@manchester.ac.uk; ORCID: 0000-0002-6080-1382},
Alexander Struck\thanks{Cluster of Excellence Image Knowledge Gestaltung at Humboldt-Universität zu Berlin; Germany; Alexander.Struck@hu-berlin.de; ORCID: 0000-0002-1173-9228}}

\maketitle

\begin{abstract}
This article summarizes motivations, organization, and activities of the Workshop on Sustainable Software for Science: Practice and Experiences (WSSSPE5.1) held in Manchester, UK in September 2017. The WSSSPE series promotes sustainable research software by positively impacting principles and best practices, careers, learning, and credit. This article discusses the Code of Conduct, idea papers, position papers, experience papers, demos, and lightning talks presented during the workshop. The main part of the article discusses the speed-blogging groups that formed during the meeting, along with the outputs of those sessions. % Finally, it discusses a survey of the workshop attendees.

% It also summarizes a set of lightning
% talks in which speakers highlighted to-the-point lessons and challenges
% pertaining to sustaining scientific software.
% The final and main contribution of the report is a summary of the
% discussions, future steps, and future organization for a set of self-organized
% working groups on topics including developing pathways to funding scientific
% software; constructing useful common metrics for crediting software
% stakeholders; identifying principles for sustainable software engineering
% design; reaching out to research software organizations around the world; and
% building communities for software sustainability. For each group, we include a
% point of contact and a landing page that can be used by those who want to join
% that group's future activities. The main challenge left by the workshop is to
% see if the groups will execute these activities that they have scheduled, and
% how the WSSSPE community can encourage this to happen.

\end{abstract}

\section*{Keywords}

sustainable software, research software engineering

\clearpage

%%%%%%%%%%%%%%%%%%%%%%%%%%%%%%%%%%%%%%%%%%%%%%%%%%%%%%%%%%%%
\section{Introduction} \label{sec:intro}
%%%%%%%%%%%%%%%%%%%%%%%%%%%%%%%%%%%%%%%%%%%%%%%%%%%%%%%%%%%%

%\todo{Dan}

In September 2017, 37 people interested in sustainable research software came together at the Working towards Sustainable Software for Science: Practice and Experiences (WSSSPE5.1, \href{http://wssspe.researchcomputing.org.uk/wssspe5-1/}{wssspe.researchcomputing.org.uk/wssspe5-1/}) meeting in Manchester, UK. This immediately preceded the Second Research Software Engineers (RSE) Conference, so that RSE attendees could also attend WSSSPE5.1.

%%%%%%%%%%%%%%%%%%%%%%%%%%%%%%%%%%%%%%%%%%%%%%%%%%%%%%%%%%%%
%\subsection{Mission and vision}\label{sec:mission}
%%%%%%%%%%%%%%%%%%%%%%%%%%%%%%%%%%%%%%%%%%%%%%%%%%%%%%%%%%%%

%\todo{should we take any of this into the intro, or drop it all?}

%\sdnote{I'm for keeping it for ``newcomers''}

%{\bf Mission.}
 WSSSPE is an international community-driven organization that promotes sustainable research software by addressing challenges related to the full lifecycle of research software through shared learning and community action.
%
% {\bf Vision.}
 It envisions a world where research software is accessible, robust, sustained, and recognized as a scholarly research product critical to the advancement of knowledge, learning, and discovery.

% {\bf Focus areas.}
 WSSSPE promotes sustainable research software by positively impacting:
 \begin{itemize}
 \item Principles and Best Practices. Promoting best practices in sustainable software
 \item Careers. Developing and supporting career paths in research software development and engineering
 \item Learning. Engaging in activities to promote peer learning and interaction
 \item Credit. Ensuring recognition of research software as an intellectual contribution equal to other research products
 \end{itemize}

% \textbf{Definitions:}
WSSSPE defines \emph{Sustainable software} as software that has the capacity to endure such that it will continue to
 be available in the future, on new platforms, meeting new needs.
 The \emph{research software lifecycle} includes:
 acquiring and assembling resources (including funding and people) into teams and communities,
 developing software,
 using software,
 recognizing contributions to and of software,
 and
 maintaining software.

Previous WSSSPE events\footnote{The first WSSSPE workshop was named ``Working towards Sustainable Software for Science: Practice and Experiences,'' which remains the meaning of the WSSSPE group, but the workshops after that were named ``Workshop on Sustainable Software for Science: Practice and Experiences.'' Together these reflect that WSSSPE is both a community and a set of workshops.},
all but the last of which were in the US, were two general, presentation-focused, one-day workshops held with the SC13 and SC14 conferences~\cite{WSSSPE1-pre-report,WSSSPE1,WSSSPE2-pre-report,WSSSPE2}, two half-day workshops (\href{http://wssspe.researchcomputing.org.uk/wssspe1-1/}{wssspe.researchcomputing.org.uk/wssspe1-1/}, \href{http://wssspe.researchcomputing.org.uk/wssspe2-1/}{wssspe.researchcomputing.org.uk/wssspe2-1/}) held with SciPy 2014 and 2015 that contained presentations about specific sustainable Python software packages, a one-and-a-half-day workshop that included teams that self-assembled and discussed focused software sustainability topics~\cite{WSSSPE3}, and a two-and-a-half-day workshop that immediately preceded the First Research Software Engineers (RSE) Conference in Manchester, England~\cite{WSSSPE4-report}.

These workshops dedicated a significant portion of their time to group discussions about problems and potential solutions in the sustainable research software space.  WSSSPE5.1 used the speed blog methodology to generate eight reports on different views of the space. The blogs were published by the UK Software Sustainability Institute (SSI) on its website.

%%%%%%%%%%%%%%%%%%%%%%%%%%%%%%%%%%%%%%%%%%%%%%%%%%%%%%%%%%%%
\subsection{Call for participation and response} \label{sec:preworkshop}
%%%%%%%%%%%%%%%%%%%%%%%%%%%%%%%%%%%%%%%%%%%%%%%%%%%%%%%%%%%%

%\sdnote{Have left out most of the reasoning behind the call, and call details. Could be long enough to warrrant a heading, but can likewise be merged with intro.}

Submissions to WSSSPE5.1 were made to two tracks, each of which accepted lightning talks (up to two pages) and papers (up to four pages).

\begin{itemize}
\item Track 1 -- ``The state of the art in sustainable research software'' - called for submissions that report and evaluate projects which created or currently create foundations, ecologies and tools for sustainable research software.
\item Track 2 -- ``Towards a sustainable future for research software'' - called for submissions that identify new or revisit existing implementable proposals for making research software sustainable.
\end{itemize}

Submissions comprised six papers (four in Track 1; two in Track 2) and eight lightning talks (six in Track 1; two in  Track 2).
Of these, four papers (three in Track 1; one in Track 2) and seven lightning talks (five in Track 1; two in Track 2) were accepted.
Similar to previous years, most submissions were accepted in order to include as many relevant inputs as possible, and to encourage their authors to share and implement their input.

All papers and lightning talks have been published as a figshare collection~\cite{WSSSPE5_1_proceedings_2017}. Slides for the given talks have been published on the WSSSPE5.1 website (\href{http://wssspe.researchcomputing.org.uk/wssspe5-1/wssspe5-1-agenda/}{wssspe.researchcomputing.org.uk/wssspe5-1/wssspe5-1-agenda/}).

%%%%%%%%%%%%%%%%%%%%%%%%%%%%%%%%%%%%%%%%%%%%%%%%%%%%%%%%%%%%
\subsection{Code of Conduct}\label{sec:CoC}
%%%%%%%%%%%%%%%%%%%%%%%%%%%%%%%%%%%%%%%%%%%%%%%%%%%%%%%%%%%%

%\todo{do we want to keep some discussion of this here?  maybe fold in into the intro?}

WSSSPE5.1 included a Code of Conduct (CoC, \href{http://wssspe.researchcomputing.org.uk/wssspe4/code-of-conduct/}{wssspe.researchcomputing.org.uk/wssspe4/code-of-conduct/}) as guidance for the community of scientists that WSSSPE
supports, including the workshop and their personal and online interactions (e.g., on
Twitter, in email lists, in Slack). This CoC is based on the
FORCE11 conference CoC~\cite{FORCE11:CoC}, which is in turn based on the Code4Lib
CoC~\cite{Code4Lib:CoC}.
In introducing the CoC, we asked participants to agree to the following main guidelines:
\begin{quote}
    WSSSPE events are community events intended for networking and collaboration
    as well as learning. We value the participation of every member of the
    community and want all attendees to have an enjoyable and fulfilling
    experience. Accordingly, all attendees are expected to show respect and
    courtesy to other attendees throughout the event and in interactions online
    associated with the event.

    The WSSSPE event organizers are dedicated to providing a harassment-free
    experience for everyone, regardless of gender, gender identity and
    expression, age, sexual orientation, disability, physical appearance,
    body size, race, ethnicity, religion (or lack thereof), technology choices,
    or other group status.

    To make clear what is expected, everyone taking part in WSSSPE events and
    discussions---speakers, helpers, organizers, and participants---is required
    to conform to the Code of Conduct.

 \end{quote}

The CoC was discussed at the beginning of WSSSPE5.1, including presenting the CoC subcommittee
and a general email address for reporting concerns or incidents, or
asking questions.

\subsection{Outline}
The remainder of this paper includes %the call for papers (\S\ref{sec:preworkshop}); the WSSSPE Code of Conduct (\S\ref{sec:CoC}) and mission and vision (\S\ref{sec:mission}); 
a set of presentations from the accepted papers and lighting talks (\S\ref{sec:presentations}); outputs from the speed-blogging groups (\S\ref{sec:speed_blogs};  thematic analysis of the resulting blogs (\S\ref{sec:speed_blog_analysis}), and discussion about attendee opinions of the meeting (\S\ref{sec:survey}), before concluding (\S\ref{sec:conclusions}).

%%%%%%%%%%%%%%%%%%%%%%%%%%%%%%%%%%%%%%%%%%%%%%%%%%%%%%%%%%%%
\section{Presentations}\label{sec:presentations}
%%%%%%%%%%%%%%%%%%%%%%%%%%%%%%%%%%%%%%%%%%%%%%%%%%%%%%%%%%%%

%\sdnote{Kept this abstract to present overview of topics covered rather than re-hashing presentations. IMHO blogs are perhaps the more important outcome. Let me know if I should also give synopses or a broad overview of topics.} \katznote{thanks for doing this - I generally like it}

WSSSPE5.1 included the presentation of four 12-minute talks based on accepted papers \cite{nangia_track_2017,haupt_track_2017,queiroz_track_2017,mulholland_track_2017},
and seven 6-minute lightning talks \cite{silva_track_2017,struck_track_2017,washbrook_track_2017,dasler_track_2017,alhozaimy_track_2017,maassen_track_2017,druskat_track_2017}.

\begin{figure}[h!]
  \centering
  \includegraphics[width=\textwidth]{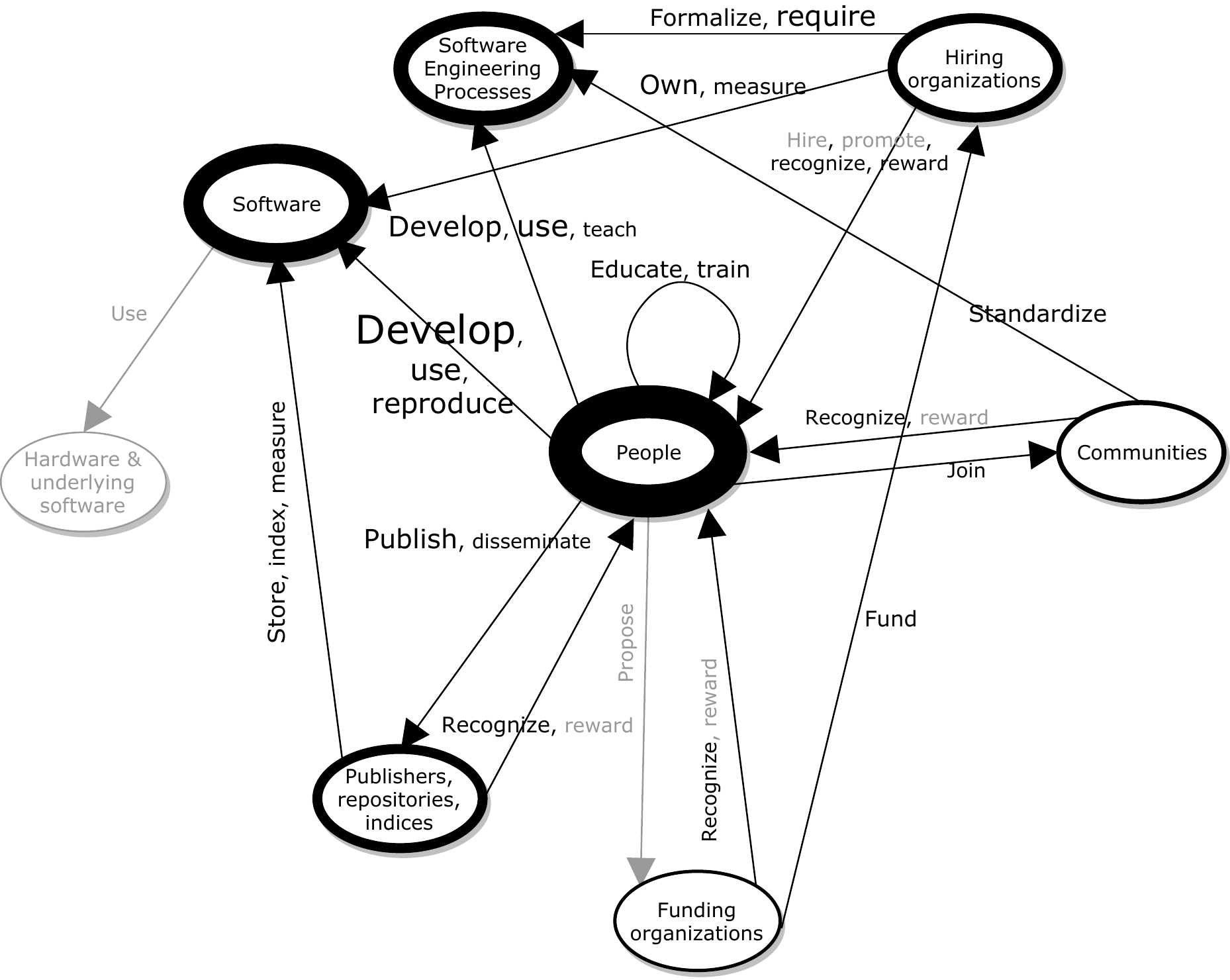}
  \caption{Research software sustainability space schematic; topic weights based on WSSSPE5.1 presentations.}
  \label{fig:schematic}
\end{figure}

Figure~\ref{fig:schematic} shows a version of a schematic of the research 
software sustainability space 
as introduced by Daniel S.~Katz \cite{katz_research_2018}.
In this version, both node outlines and edge labels are weighted to show the distribution of
their references from the workshop presentations \cite{druskat_activity}.

\begin{figure}[h!]
  \includegraphics[width=\textwidth]{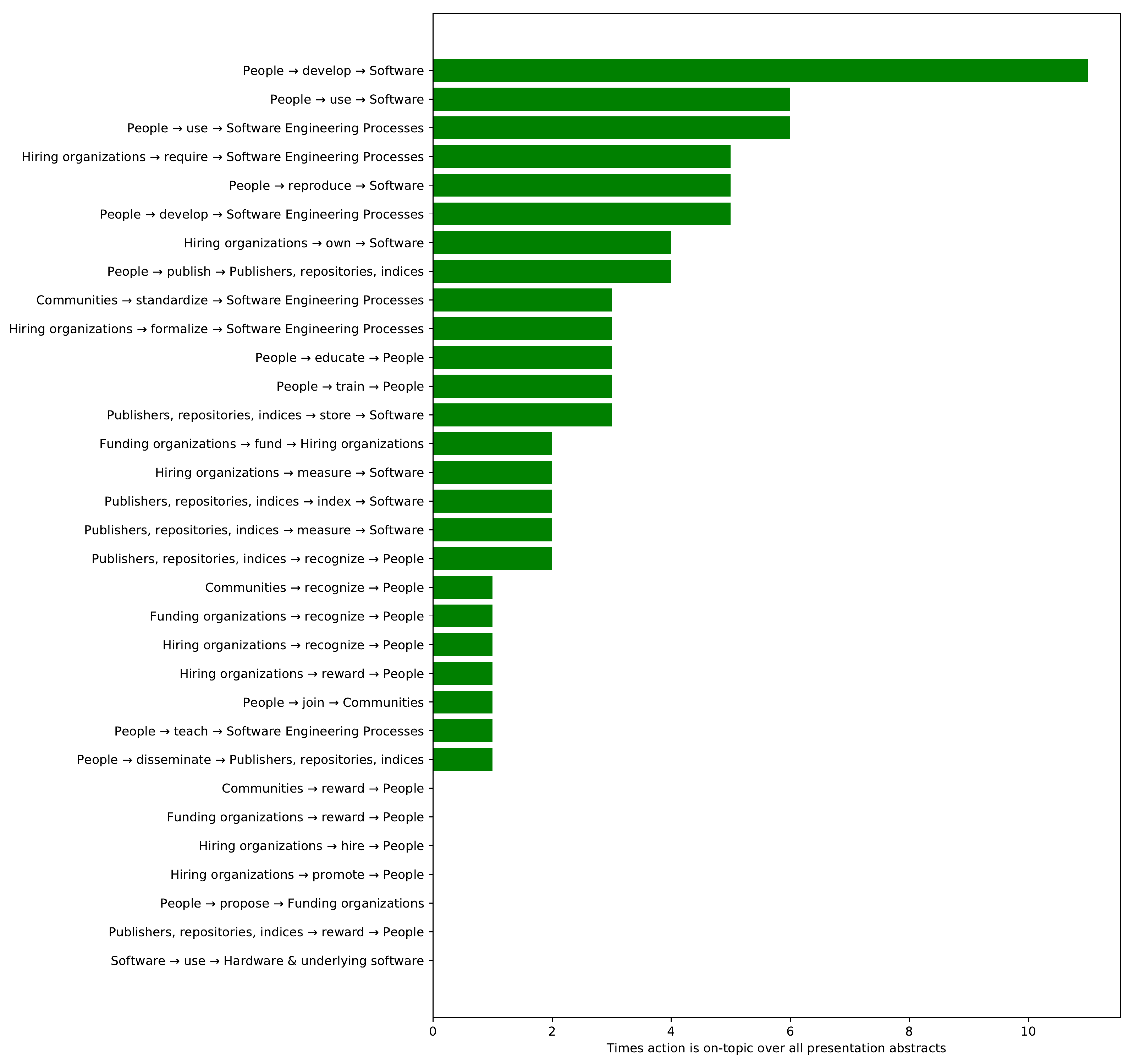}
  \caption{Distribution of action coverage over presentations.}
  \label{fig:actions}
\end{figure}

Figure~\ref{fig:actions} shows the distribution of actions over all presentations, where an action is a labeled edge from a node in the sustainability schematic (Figure~\ref{fig:schematic}) to another node. The source node represents the actor, and the target node the actee of the respective action.
The weights and action distribution show a clear overall primary focus of presentations on software development (\textit{People develop Software}). More generally, the people involved in research software take prominent focus as well as software engineering principles, and research software itself. However, across all presentations most actors, actees, and actions within the space have been the topic of a presentation as actor or actee, with the exception of hardware and underlying software. Future workshops could take care to address this topic specifically in their calls for submissions in order to close gaps in research, discussion and progress.

%This coverage of topics continues in the speed blogs, which were partly based on presentations, as discussed in a presentation by Daniel S.~Katz \cite{katz_research_2018}.

Table~\ref{tab:topics} shows the distribution of combined actor and actee reference from the
research software sustainability space over workshop
presentations.

\begin{table}[h!]
\centering
\footnotesize
\begin{tabular}{ccccccccc}
%	\hline
    	\rot{\shortstack[l]{Presentation}} & \rot{Communities} & \rot{\shortstack[l]{Funding \\ organizations}} & \rot{\shortstack[l]{Hardware \& \\ underlying \\ software}} & \rot{\shortstack[l]{Hiring \\ organizations}} & \rot{People} & \rot{\shortstack[l]{Publishers, \\ repositories, \\ indices}} & \rot{Software} & \rot{\shortstack[l]{Software \\ engineering \\ processes}} \\
        \hline
\cite{nangia_track_2017} &                                 &                                          &                                                    &                                         & •                          &                                                     & •                            &                                                   \\ 
\hline
\cite{haupt_track_2017} & •                               &                                          &                                                    & •                                       & •                          &                                                     & •                            & •                                                 \\ 
\hline
\cite{queiroz_track_2017} &                                 &                                          &                                                    &                                         & •                          &                                                     & •                            & •                                                 \\ 
\hline
\cite{mulholland_track_2017} &                                 &                                          &                                                    & •                                       & •                          &                                                     & •                            & •                                                 \\ 
\hline
\cite{silva_track_2017} &                                 & •                                        &                                                    & •                                       & •                          &                                                     & •                            &                                                   \\ 
\hline
\cite{struck_track_2017} &                                 & •                                        &                                                    & •                                       & •                          &                                                     & •                            & •                                                 \\ 
\hline
\cite{washbrook_track_2017} &                                 &                                          &                                                    & •                                       & •                          &                                                     & •                            & •                                                 \\ 
\hline
\cite{dasler_track_2017} &                                 &                                          &                                                    & •                                       &                            & •                                                   & •                            &                                                   \\ 
\hline
\cite{alhozaimy_track_2017} &                                 &                                          &                                                    &                                         & •                          & •                                                   & •                            & •                                                 \\ 
\hline
\cite{maassen_track_2017} & •                               &                                          &                                                    & •                                       & •                          & •                                                   & •                            & •                                                 \\ 
\hline
\cite{druskat_track_2017} & •                               & •                                        &                                                    & •                                       & •                          & •                                                   & •                            & •                     \\
\hline
\end{tabular}
\caption{Distribution of topics from the research software sustainability
space over workshop presentations for actors and actees (combined).}
\label{tab:topics}
\end{table}

%%%%%%%%%%%%%%%%%%%%%%%%%%%%%%%%%%%%%%%%%%%%%%%%%%%%%%%%%%%%
\section{Speed blogs} \label{sec:speed_blogs}
%%%%%%%%%%%%%%%%%%%%%%%%%%%%%%%%%%%%%%%%%%%%%%%%%%%%%%%%%%%%

%\todo{Alexander}

%\note{explain the process of speed blogging}\\
% https://www.software.ac.uk/term/speed-blogging
The Software Sustainability Institute (SSI) discusses speed blogging at \href{https://www.software.ac.uk/term/speed-blogging}{www.software.ac.uk/term/speed-blogging}. They provide helpful approaches and pointers to some valuable resources on academic writing.
The goal of speed blogging is to preserve as much content and context as possible from working groups, and to publish results in an easily digestible form. The output is usually published as blog posts, but other formats, such as an audio podcast or an informative table, could also serve the same purpose. During the process it is important to reserve at least of half the available working group time for creating the desired output format. The recommended best practice is to work collaboratively on such a document and review assigned chapters.
The SSI says,
\begin{quotation}
Speed blogs can be completed and be publication ready during the allotted time at an event or they can be near complete [at] the event and then tidied up and made ready for publication soon after the event. Thus they are time bound, time sensitive and are a fixed task that does not represent an ongoing commitment.
\end{quotation}

%\note{summarize the speed blogs themselves, with citations/pointers to them on the SSI page}\\
The rest of this section summarizes the WSSSPE5.1 blog posts, which are accessible at \href{http://wssspe.researchcomputing.org.uk/wssspe5-1/}{wssspe.researchcomputing.org.uk/wssspe5-1/}. Indented text indicated quotations from the blogs.

\begin{itemize}
\item The Research Software Project Manager (\href{https://www.software.ac.uk/blog/2017-12-04-research-software-project-manager}{www.software.ac.uk/blog/2017-12-04-research-software-project-manager})
\begin{quote}
For many, the role of research software project manager (RSPM) may be an accidental calling. The career path for this role isn’t well-established, and research software development in academia may itself be something of a haphazard, nigh-accidental byproduct of conducting domain research. Individuals approaching this role may have little to no wider industry experience, instead approaching the project manager role from research or research software engineering.
\end{quote}
The authors present the challenges for the RSPM role, noting that is inherently different from similar positions in the industry. The focus of work is on enabling research as an iterative process, prone to many changes along the way. While many academics excel at winning grants and pushing the scientific frontiers, only a few are trained in software project management. And those who are may focus on technical aspects, discarding parts of the overall picture. This raises many questions, as documented in the blog post, and offers answers to a few of them, include strengthening the role of the RSPM and considering positions dedicated to project management. Such a delegation of tasks would smooth the process of reporting to administration and free up necessary resources for actual research.

\item Looking for software use in research (\href{https://www.software.ac.uk/blog/2017-12-05-looking-software-use-research}
{www.software.ac.uk/blog/2017-12-05-looking-software-use-research}
\begin{quote}
Nowadays, software is used in most research. But how the software is created, used, and what it depends on are not well understood questions. The importance of such knowledge varies based on the motivation of the reader. On one side, we could be interested in the impact of the software, how many times it has been used and by who. This type of analysis could come, for example, from funding bodies and organisations to reward the creation of something and help its sustainability, from institutions who hire people behind that software, or from the software authors to get an understanding of the needs of their users or simply to get credit for their work. Another motivation may be trying to understand the research being carried out with a particular software or set of tools either for purely academic purposes (e.g., by historians and scholars of science) or with a commercial perspective (such as by intellectual property teams from universities for the monetisation of the software). %lx% should be shortened
\end{quote}
The blog post reports on methods for finding software. Researchers usually consult their favorite search engines and programming language package archives. In the future, Current Research Information Systems (CRIS) or repositories maintained by funders may hold information about software developed during research. Software as a research output is increasingly registered with DOIs \cite{Fenner2018DOI} but identifying software in written text still has its challenges \cite{Li2017R}. Repositories such as \href{https://www.bioconductor.org/}{Bioconductor} offer some domain-specific overview of available tools. General purpose repositories, such as \href{http://github.com}{GitHub.com}, track usage as in forks on their platforms, whereas services like \href{http://dpesy.org}{Depsy.org} and \href{https://libraries.io/}{libraries.io} provide aggregated software usage information from many sources. %lx% put links in there
Software, considered by some to be a subset of research data, could also be identified resolving PIDs like DOIs, minted by services like \href{http://datacite.org}{DataCite.org} that point to published data and software. 
%Efforts describing relevant research software repositories in a meta catalog have been undertaken by, e.g., \href{http://re3data.org}{re3data.org}, \katznote{re3data is really a catalog of repositories that meet some guidelines, so I'm not sure if makes sense to discuss it here, or maybe I'm misunderstanding} but currently fail to provide consistent indexing and comprehensive search functionality. \LXnote{It's not a discussion but rather mention of a failed effort to make software findable} \katznote{Are you saying that an intent of the re3data.org project was to make software findable, or that after re3data.org started for other purposes, it had an effort to make software findable?} \LXnote{by definition of the funding agency (DFG), research software is a subset of research data, therefore the intent of re3data.org is to make research data AND software findable ... by cataloging repositories that hold data and/or software. The resulting platform has a lot of room for improvement but is mentioned here as a different way of pointing researchers to software. If that is too confusing for the reader, let me know and I will remove the sentence.} \katznote{I see what you are saying, but it's very indirect and confusing to me without a longer explanation, such as you have provided in the comments - maybe remove this?}

\item Towards Reproducibility in Research Software (\href{https://www.software.ac.uk/blog/2017-12-06-towards-reproducibility-research-software}{www.software.ac.uk/blog/2017-12-06-towards-reproducibility-research-software})
\begin{quote}
Ensuring reproducibility of research has been identified as one of the challenges in scientific research. While reproducibility of results is a concern in all fields of science, the emphasis of this group is in the area of computer software reuse and the reproduction of results. The availability of complete descriptions, ideally including program source code, documentation and archives of all necessary components and input datasets would be a major step to resolving research reproducibility concerns.
\end{quote}
A short review of relevant services and platforms is provided by the authors. It includes concepts from software engineering, e.g., version control, being introduced in programmes such as \href{https://software-carpentry.org}{Software Carpentry}.  %lx% link
Platforms and tools, such as \href{https://www.overleaf.com}{Overleaf} or \href{http://jupyter.org}{Jupyter notebooks}, ease collaborative review of research and reproducibility of code respectively. While the uptake of such best practices is slow and the lack of documentation sometimes hinders reproducibility, we may see more of them in the future, especially if incentives are set right.

\item Why research software engineers should have permanent contracts (\href{https://www.software.ac.uk/blog/2017-12-11-why-research-software-engineers-should-have-permanent-contracts}{www.software.ac.uk/blog/2017-12-11-why-research-software-engineers-should-have-permanent-contracts})

\noindent At present, only a few research institutions employ research software engineers and make their resources available to the whole organization. This blog post discusses some success stories motivating long term contracts for RSEs. In general ``better software [leads to] better research'' (\href{https://software.ac.uk/}{SSI}).
\begin{quote}
Higher quality output is on every university's wish list, as it leads to a potential increase in QR funding (\href{http://www.hefce.ac.uk/rsrch/funding/mainstream/}{www.hefce.ac.uk/rsrch/funding/mainstream/}), while reducing the reputational risk associated with substandard research practices.
Reproducibility is notoriously difficult to achieve and RSEs are an essential part of enabling this [\ldots] leading to higher citations and greater research impact.
\end{quote}
Skilled RSE are sought after and organizations get what they pay for. The ``stability and the potential for some form of career progression'' that permanent contracts offer should be a consideration when hiring such staff. And the costs do not necessarily reduce the university's baseline funds as research projects are encouraged to ``cost in'' consulting by centrally pooled RSEs. Long-term benefits for organizations are the spreading of best practices in software development and an institutional memory provided by the central RSE group who have worked in many projects. 
\begin{quote}
An RSE team that is permanently employed can be truly agile. Recruiting an expert for a short period of time in an academic institution is virtually impossible. As long as we rely on fixed-term contracts for RSEs, a lot of important work will fail to be done, and funds will not be spent as effectively as they could be.
\end{quote}

\item A standard format for CITATION files (\href{https://www.software.ac.uk/blog/2017-12-12-standard-format-citation-files}{www.software.ac.uk/blog/2017-12-12-standard-format-citation-files})
\begin{quote}
The citation of research software has a number of purposes, most importantly attribution and credit, but also the provision of impact metrics for funding proposals, job interviews, etc. Stringent software citation practices [\ldots] therefore include the citation of a software version itself, rather than a paper about the software. Direct software citation also enables reproducibility of research results as the exact version can be retrieved from the citation. 
\end{quote}
While a diverse range of citation recommendations exist for software projects, many projects still proceed without a proper citation. In order to unify the various existing  procedures, the authors proposed the inclusion of a CITATION file in a software repository, holding standardized information, e.g., ORCID to enable linking authors with their respective ID from \href{https://orcid.org}{orcid.org}. It shall be easy to read and write in order to enable fast and straightforward creation of citation information. For this purpose, the site \href{https://research-software.org/citation/}{research-software.org/citation/} was created and is actively maintained to cover all aspects of research software citation. Widespread use of the Citation File Format (CFF) may enable transitive credit information provision and better impact metrics next to reusability by other actors.

\item Overcoming barriers to adopting software best practices in research (\href{https://www.software.ac.uk/blog/2017-12-07-overcoming-barriers-adopting-software-best-practices-research}{www.software.ac.uk/blog/2017-12-07-overcoming-barriers-adopting-software-best-practices-research})

\noindent Research careers create a wide spectrum of skill levels (compared to corporate environments). Training a group of researchers on a new topic is challenging, given the lack of a peer group at some research frontiers or missing institutional support for training and the significant upfront cost involved in learning e.\,g., best development practices. Researchers who mostly work on code in isolation rarely have an opportunity to have code reviews (with a notable exception as described in \href{https://www.software.ac.uk/blog/2018-05-18-code-review-academia}{www.software.ac.uk/blog/2018-05-18-code-review-academia}), need to learn on their own, and are often not recognized for writing code. Paper publication in a ``high-impact peer-reviewed journal'' still pushes careers. A way to overcome barriers is to introduce, e.g., ``industry standards'' in an manner adapted to research environments. The blog post discusses how the ``SCRUM'' approach may not work, but ``agile'' or a ``maturity model'' approach may better fit the current research practices.
\begin{quote}
As a general concept: start small and then go as far as necessary. Reaching for the perfect software development approach is intimidating and overwhelming, and it is not the task of a researcher nor necessary for most research projects. A maturity model can help researchers identify where they are and where they should be [\ldots]. Restricting the use of tech jargon to a minimum and offering explanations where necessary can help, too.
\end{quote}

\item Encouraging good software development practice in research teams (\href{https://www.software.ac.uk/blog/2017-12-13-encouraging-good-software-development-practice-research-teams}{www.software.ac.uk/blog/2017-12-13-encouraging-good-software-development-practice-research-teams}

\noindent ``Training in good software development skills is vital for the uptake and maintenance of good practice in a research community.'' This requires that resources, tools, and sometimes management support the availability of, and that the trainee be motivated. The latter could be achieved when the ``connection between good practice and increased publication rate and quality, as well as funding availability'' are demonstrated. A ``consistent approach to software development [\ldots] can benefit collaboration and the longevity of projects.'' Success story from the DLR (\href{https://www.dlr.de}{www.dlr.de}) and the EMBL (\href{https://www.embl.de/}{www.embl.de}) are reported in the blog post, which goes into detail on how to build a community and keep it alive. The authors propose the concept of ``inner source,'' where open source principles are utilized collaboratively inside an organization to develop ``common scientific frameworks.'' GitLab is praised as a beneficial tool that eases adoption of good practices. The authors recognize that rollout of training, tools, and services should happen iteratively and is always highly dependent on the particular environment. Therefore they are interested to learn from others and encourage feedback via e-mail.

\item Overcoming Entry Barriers to Motivate Better Practice in Research Software Engineering (\href{https://www.software.ac.uk/blog/2017-12-14-overcoming-entry-barriers-motivate-better-practice-research-software-engineering}{www.software.ac.uk/blog/2017-12-14-overcoming-entry-barriers-motivate-better-practice-research-software-engineering})
\begin{quote}
What can be termed as ``coding'' is a subset of wider software engineering practices such as version control, continuous integration and good software design. Coding is prevalent in academia but practices that allow sustainable software to be produced are frequently overlooked.  Motivating the uptake of the approaches, methods and tools, and highlighting the benefit they deliver, by engaging with researchers who develop software is the first step in spreading best practice in our community.
\end{quote}
The authors point out the benefit of using online systems such as GitLab to reduce entry barriers and motivating the use of, e.g., version control and continuous integration (CI). Other software engineering principles such as pair-programming and code review are also encouraged early in (graduate) students training in order to demonstrate the benefits for future use and bridging gaps between disciplines. If these research software management practices become a requirement in grants application and reporting, widespread deployment is inevitable. Increasing recognition of software as a valid research output could provide further motivation, along with better reproducibility and reduced duplication of effort.

\end{itemize}

%%%%%%%%%%%%%%%%%%%%%%%%%%%%%%%%%%%%%%%%%%%%%%%%%%%%%%%%%%%%
\section{Analyzing the speed blogs} \label{sec:speed_blog_analysis}
%%%%%%%%%%%%%%%%%%%%%%%%%%%%%%%%%%%%%%%%%%%%%%%%%%%%%%%%%%%%

%\sdnote{Alexander, all, would it be interesting to look into topic distribution for the speed blogs as well? Dan has had a slide on it at another talk (referenced in section above), but could possibly be done in more detail.}
%\katznote{I was thinking that might be part of what Caroline and Rob would do in this section.}

%\todo{Caroline and Rob}

%\note{explain the results?}

The speed blog topics were determined using an unconference format---where participants chose what to cover as a group---and therefore provided a snapshot of issues that were particularly timely and relevant to the WSSSPE community. Participants were free to choose which speed blogging group to join, depending on their own particular interests and goals. In this section, we treat the speed blogs as qualitative data, which we systematically analyse to determine the prevalence of particular themes.

\subsection{Method} 
We used a hybrid thematic/framework analytic approach, where we used the schematic of the research software sustainability space shown in Figure~\ref{fig:schematic} as a starting point for our analysis, and then refined this as we familiarized ourselves with the data. The schematic nodes formed the categories: \emph{funders} (`funding organizations' in the schematic); \emph{employers} (`hiring organizations'); \emph{publishers, repositories, indices}; \emph{research software} (`software'); \emph{software engineering processes}; \emph{communities}. The edges provided the categories: \emph{reward \& recognition} (`recognize, reward'); \emph{training}; \emph{standardization} (`standardize'); \emph{reproducibility} (`reproduce'). Based on a bottom-up analysis of the data, we broke the schematic `people' node down further into \emph{users}, \emph{research software engineers} and \emph{researchers}, and added a further category of \emph{software infrastructure}.

Authors Jay and Haines coded the blogposts independently, recording for each blogpost whether or not the theme was present. This process resulted in agreement of 79\%. Disagreements were then resolved via discussion. The original data set (blogposts and individual and joint coding scores) are available for further analysis~\cite{data:speed-blog-analysis}.

\subsection{Results}
\begin{figure}[h!]
  \includegraphics[width=\textwidth]{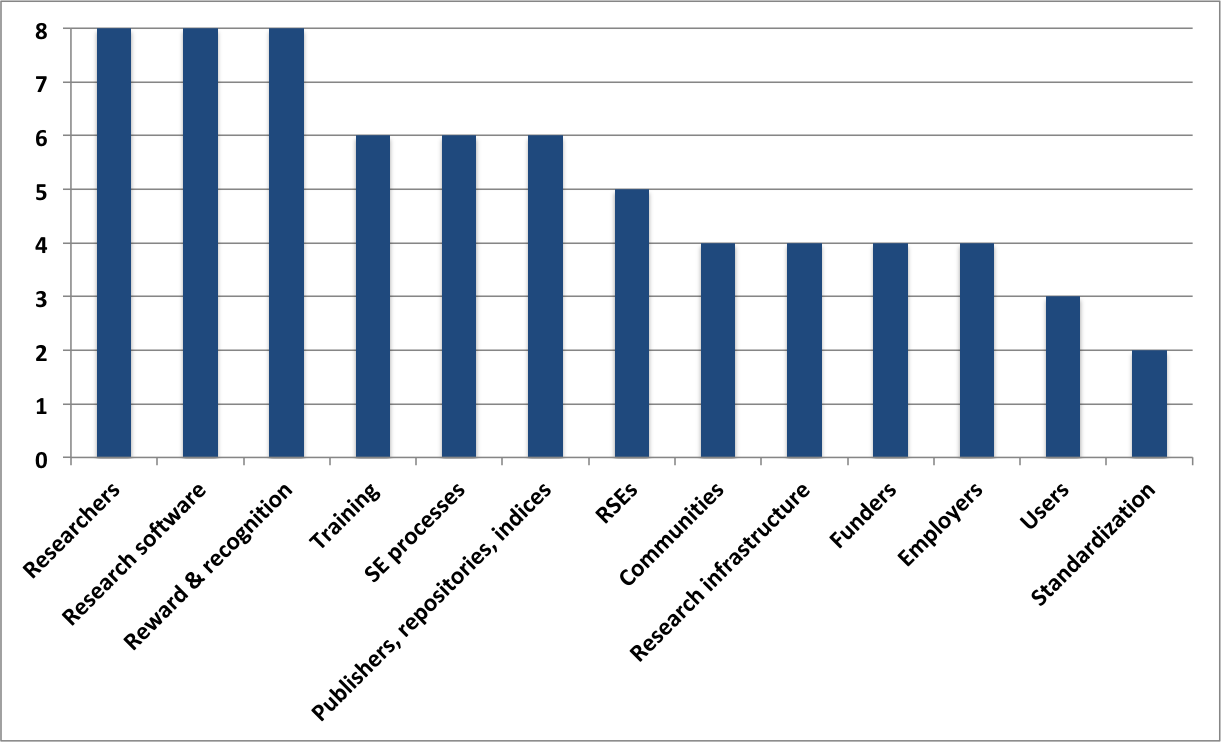}
  \caption{The number of speed blogs mentioning each of the themes.}
  \label{fig:speedblogs}
\end{figure}

All of the eight blogs mentioned research software, and researchers (i.e., domain specialists rather than RSEs). The blogs also all mentioned reward and recognition, either for software itself, or for people writing software. The prevalence of all of the categories across the blog posts can be seen in Figure~\ref{fig:speedblogs}. During the discussion of the speed blog topics it was decided that there seemed to be sufficient writing already on citation and credit, so while this topic fit under the WSSSPE umbrella, it was not suggested as a topic for speed blogs. The fact that it was mentioned in all the blogs indicates that it is still an important theme for the community.

\section{Attendee survey} \label{sec:survey}
%%%%%%%%%%%%%%%%%%%%%%%%%%%%%%%%%%%%%%%%%%%%%%%%%%%%%%%%%%%%

At the end of WSSSPE5.1, participants were asked to complete an online survey, and 35 of the 37 did \cite{druskat_survey}.
The answers indicate that respondents were very satisfied with the workshop in general. Similarly to WSSSPE4\cite{WSSSPE4-report}, the mixture of topics and speakers as well as the proportions of longer presentations and lightning talks was deemed well-balanced, while some responses indicate that more time for discussion would be helpful, which may include an extension of the workshop length. The majority of respondents (73\%) found the discussion/speed blogging session more valuable than the presentation session, which seems to support the hands-on approach of the workshop format. In general, respondents found the purpose of the workshop sufficiently clear, and generally agree that its purpose was fulfilled.

Similar to the results of the WSSSPE4 attendee survey \cite{WSSSPE4-report}, the results of the current survey indicate that future WSSSPE events should include a review of basic concepts and terminology: 11\% of respondents somewhat disagree that the concept of ``software sustainability'' as presented at the workshop is well-defined. Similarly, the overall vision and progress of WSSSPE as well as the organization of concrete actions between WSSSPE workshops should be defined and communicated more strongly. Additional topics to explore and include in future events were case studies (also mentioned in the WSSSPE4 survey results); the organization of Research Software Engineering teams; funding for software fellowships; measurement and composition of communities, as well as tools, services and user engagement practices; impact metrics for sustainable software; and guidance on making the case for sustainability.

Some participants found the length of the lightning talk (six minutes) to be odd. They suggested that their length should be reduced to more typical talking times for this type of presentation (two-to-three minutes), and that the time saved as a result should be used for keynotes at the beginning of the workshop. Alternative suggestions were that all presentations should be of equal length (12-15 minutes). Other suggestions included making the workshop available virtually or provide recordings, and repeatedly, to make it longer than one day.

%%%%%%%%%%%%%%%%%%%%%%%%%%%%%%%%%%%%%%%%%%%%%%%%%%%%%%%%%%%%
\section{Conclusions} \label{sec:conclusions}
%%%%%%%%%%%%%%%%%%%%%%%%%%%%%%%%%%%%%%%%%%%%%%%%%%%%%%%%%%%%

%\todo{more needed?}
In summary, the community recognizes the need for improving software engineering practices and takes action. A set of initial success stories have been reported from a handful of organizations. Recognition for work, including via citations, remains a topic of interest. 

Based on the presentations and working groups at WSSSPE, we have presented a set of topics and mapped the WSSSPE5.1 presentations and speed blogs onto these topics. The presentations and speed blogs cover most of the topics and do not have subjects that are not in the list, indicating this that topics (or themes, a more fine-grained mapping of the space as discussed in Section~\ref{sec:speed_blog_analysis}) are a good representation of the space.

%The discussions and hands-on sessions were the most valued parts of the workshop by the participants, who requested that the overall duration of the workshop should be lengthened. \LXnote{thinks that this could work as conclusions.} 

%%%%%%%%%%%%%%%%%%%%%%%%%%%%%%%%%%%%%%%%%%%%%%%%%%%%%%%%%%%%
%\section*{Acknowledgments} \label{sec:acks}
%%%%%%%%%%%%%%%%%%%%%%%%%%%%%%%%%%%%%%%%%%%%%%%%%%%%%%%%%%%%

%\katznote{feel free to add stuff here}

\newpage
\bibliographystyle{vancouver}

\bibliography{wssspe.bib}
\end{document}